\documentclass[pra,twocolumn,twoside,showpacs]{revtex4}

\usepackage{ae}
\usepackage{graphicx}
\usepackage{amsmath,bm}

\newcommand{\eq}[1]{\begin{equation}#1\end{equation}}
\newcommand{\eqmulti}[1]{\begin{equation}\begin{split}#1\end{split}\end{equation}}

\newcommand{\ket}[1]{\ensuremath{\,|{#1}\rangle}}
\newcommand{\braket}[2]{\ensuremath{\langle{#1}|{#2}\rangle}}
\newcommand{\matrixe}[3]{\ensuremath{\langle{#1}|\,{#2}\,|{#3}\rangle}}

\newcommand{\op}[1]{\ensuremath{\bm{#1}}}
\newcommand{\dd}{\ensuremath{\mathrm{d}}}

\newcommand{\AO}{\ensuremath{\op{A}}}
\newcommand{\HO}{\ensuremath{\op{H}}}

\newcommand{\xV}{\ensuremath{\vec{x}}}

\newcommand{\pOV}{\ensuremath{\vec{\op{p}}}}
\newcommand{\qOV}{\ensuremath{\vec{\op{q}}}}
\newcommand{\rOV}{\ensuremath{\vec{\op{r}}}}

\newcommand{\xOV}{\ensuremath{\vec{\op{x}}}}

\newcommand{\EC}{\ensuremath{\mathcal{E}}}
\newcommand{\FC}{\ensuremath{\mathcal{F}}}

\newcommand{\nablaV}{\ensuremath{\vec{\nabla}}}

\newcommand{\PiO}{\ensuremath{\op{\Pi}}}

\newcommand{\crit}{\ensuremath{\text{cr}}}
\newcommand{\sep}{\ensuremath{\text{sep}}}
\newcommand{\B}{\ensuremath{\text{B}}}
\newcommand{\F}{\ensuremath{\text{F}}}
\newcommand{\BB}{\ensuremath{\text{BB}}}
\newcommand{\FF}{\ensuremath{\text{FF}}}
\newcommand{\BF}{\ensuremath{\text{BF}}}

\newcommand{\nB}{\ensuremath{n_{\text{B}}}}
\newcommand{\nF}{\ensuremath{n_{\text{F}}}}

\newcommand{\PhiB}{\ensuremath{\Phi_{\text{B}}}}

\newcommand{\aB}{\ensuremath{a_{\text{B}}}}
\newcommand{\aBF}{\ensuremath{a_{\text{BF}}}}

\newcommand{\mB}{\ensuremath{m_{\text{B}}}}
\newcommand{\mF}{\ensuremath{m_{\text{F}}}}
\newcommand{\mBF}{\ensuremath{m_{\text{BF}}}}

\begin{document}

\title{Structure and stability of trapped atomic boson-fermion mixtures}

\author{Robert Roth}
\email{robert.roth@physics.ox.ac.uk}

\affiliation{Clarendon Laboratory, University of Oxford,
  Parks Road, Oxford OX1 3PU, United Kingdom}

\date{\today}

\begin{abstract}
The structure of trapped binary mixtures of bosonic and fermionic
atoms at zero temperature is studied using a modified Gross-Pitaevskii
equation for the bosons which self-consistently includes the
mean-field interaction generated by the fermionic cloud. The density
of the fermionic component is described within the Thomas-Fermi
approximation. The influence of the boson-boson and the boson-fermion
$s$-wave interaction on the density profiles and the stability of the
mixture is investigated systematically. Critical particle numbers for
the mean-field collapse caused either by attractive boson-boson or by
attractive boson-fermion interactions and for the onset of spatial
component separation are discussed. It is shown that the interplay
between the boson-boson and the boson-fermion interaction generates a
rich and complex phase diagram. Finally, the specific properties and
prospects of the boson-fermion mixtures available in present
experiments are addressed.

\end{abstract}


\maketitle

\section{Introduction}

Since the first realizations of Bose-Einstein condensation in trapped
dilute atomic gases \cite{Vare99} this lively field of research has
generated an amount of impressive experimental results illuminating
basic quantum phenomena. Besides the studies using bosonic atoms first
results were obtained on the cooling of fermionic atoms to a
temperature regime where quantum effects dominate the properties of
the gas \cite{DeJi99,GrGe01}. One of the exciting prospects is the
observation of a BCS transition of these degenerate Fermi gases to a
superfluid state.

Several successful attempts to trap and cool mixtures of a bosonic and a
fermionic species were reported. Quantum degeneracy was first reached with
mixtures of bosonic ${}^{7}\text{Li}$ and fermionic ${}^{6}\text{Li}$
atoms \cite{TrSt01,ScKh01,ScFe01}. More recently, experiments to cool
mixtures of different atomic elements, i.e. ${}^{23}\text{Na}$ and
${}^{6}$Li \cite{HaSt01} as well as ${}^{87}\text{Rb}$ and ${}^{40}$K
\cite{RoRi02,GoPa02}, to ultra-low temperatures succeeded. These
boson-fermion mixtures offer unique possibilities to study the effects of
quantum statistics directly. Besides their interesting physics these
systems can be used as an efficient tool to produce a degenerate
single-component (spin-polarized) Fermi gas. Direct evaporative cooling is
not applicable in a gas of spin-polarized fermions since $s$-wave
interactions are suppressed by the Pauli principle. So the Bose gas, which
can be cooled evaporatively, is used as a coolant.  The fermionic cloud
remains in thermal equilibrium with the cold Bose gas through
boson-fermion interactions in the region where both clouds overlap. In
this way the fermionic species is cooled sympathetically \cite{AbVe97}.
Evidently, the shape of the two density distributions as well as their
stability are extremely important for the success of the sympathetic
cooling scheme.

In this paper I want to investigate systematically the influence of
inter-atomic interactions on the structure and stability of binary
boson-fermion mixtures at zero temperature. In extension to the work
presented in \cite{RoFe02} the influence of boson-boson and boson-fermion
interactions on the structure of the spatial density distributions and the
stability of the system against collapse and phase separation is
investigated. All combinations of attractive and repulsive interactions are
considered and the implications for present experiments are discussed.

Previous studies of these mixtures mainly rely on the Thomas-Fermi
approximation to treat the bosonic component
\cite{Molm98,AkVi02}. This restricts the range of applicability
significantly; the Thomas-Fermi approximation for the bosonic
component is valid only for sufficiently strong repulsive boson-boson
interactions and for large boson numbers. Other authors use a
variational model, where the bosonic density profile is parameterized
by a Gaussian \cite{MiSu01,AmMi99}. In many cases this does not
provide enough flexibility to describe the density profiles accurately.

To avoid these restrictions from the outset and obtain a model that
can be applied for all values of the scattering lengths and for all
relevant particle numbers the density of the bosonic component will be
obtained from the self-consistent solution of a modified
Gross-Pitaevskii equation which is coupled to the fermionic density
distribution through the boson-fermion interaction
\cite{RoFe02,NyMo99}. The fermionic component is treated in
Thomas-Fermi approximation which actually is a very good approximation
to describe a degenerate Fermi gas \cite{RoFe01c,RoFe01a}.

The paper is structured as follows: In Section \ref{sec:mod} the coupled
set of differential equations that describes the bosonic and fermionic
density distributions is derived and its numerical treatment is discussed.
Using this universal tool the generic properties of binary mixtures are
investigated in detail in Sections \ref{sec:BF-} and \ref{sec:BF+}.
Depending on the particle numbers, the boson-boson scattering length, and
the boson-fermion scattering length the instability of the mixture against
mean-field collapse and spatial component separation is discussed. 
Finally, in Section \ref{sec:exp} the consequences for the particular
boson-fermion mixtures available in present experiments are summarized.
\vfil

\section{Model for dilute boson-fermion mixtures}
\label{sec:mod}

We aim at the description of ground state properties of a dilute
mixture of $N_{\B}$ bosonic and $N_{\F}$ fermionic atoms in an
external trapping potential at zero temperature. The atoms can be
considered as inert interacting bosonic or fermionic particles,
because the energy scales involved are sufficiently small, i.e.,
internal excitations of the atoms are not relevant.

\subsection{Mean-field approximation and effective Hamiltonian}

A simple starting point for the treatment of the quantum mechanical
many-body problem of $N=N_{\B}+N_{\F}$ interacting bosons and fermions
is the mean-field approximation. The many-body state that describes
the binary mixture can be written as a direct product of a bosonic
$N_{\B}$-body state $\ket{\Phi_{\B}}$ and a fermionic $N_{\F}$-body
state $\ket{\Psi_{\F}}$
\eq{ \label{eq:mod_state}
  \ket{\Psi}
  = \ket{\Phi_{\B}} \otimes \ket{\Psi_{\F}} .
}
Within the mean-field approximation each of these states is described
by a symmetrized or antisymmetrized product of single-particle
states. At zero temperature the bosonic state is a direct product of
$N_{\B}$ identical single-particle states $\ket{\phi_{\B}}$
\eq{ \label{eq:mod_state_bosons}
  \ket{\Phi_{\B}}
  = \ket{\phi_{\B}}\otimes\cdots\otimes\ket{\phi_{\B}}
}
which is symmetric per se. The fermionic state necessarily is a Slater
determinant, i.e. an antisymmetrized product of different
single-particle states $\ket{\psi_1},...,\ket{\psi_{N_{\F}}}$,
\eq{ \label{eq:mod_state_fermions}
  \ket{\Psi_{\F}}
  = \AO\big(\ket{\psi_1}\otimes\ket{\psi_2}\otimes\cdots
    \otimes\ket{\psi_{N_{\F}}}\big) ,
}
where $\AO$ is the antisymmetrization operator.

The mean-field picture is a reasonable starting point as long as
correlations induced by the two-body interaction between the
constituents are weak. The two-body potential that describes the
interaction between the atoms consists of an attractive part at larger
particle distances and a strong repulsion at short
distances. Especially the short-range repulsion, the so called core,
induces strong correlations between the atoms: The probability to find
two particles within the range of the core is nearly zero, i.e., the
two-body density shows a pronounced hole for particle distances
smaller than the radius of the core. Correlations of this kind cannot
be described in mean-field approximation.

In order to use the mean-field approximation for the description of
the interacting system one has to replace the full atom-atom potential
by a suitable effective interaction that contains the relevant
physical properties of the the original potential \cite{RoFe01a}.
Since the gases we are considering are extremely dilute and cold the
particles do not resolve the detailed shape of the
potential. Therefore one may replace the interatomic potential by a
simple contact interaction for all partial waves. The strengths of the
contact terms are adjusted such that the expectation values of the
contact interaction in two-body product states reproduce the two-body
energy spectrum of the original atom-atom potential. This leads to a
closed form of the effective contact interaction for all partial waves
which involves only the scattering lengths of the original potential
as discussed in detail in Ref. \cite{RoFe01a}.

Using the effective contact interaction one can set up the Hamiltonian
of a binary boson-fermion mixture. It consist of a part $\HO_{\B}$
that involves only the bosonic component, a part $\HO_{\F}$ which
describes only the fermionic component, and a part $\HO_{\BF}$ which
represents the interaction between the two species
\eq{ \label{eq:mod_hamiltonian}
  \HO 
  = \HO_{\B} + \HO_{\F} + \HO_{\BF} .
}
The bosonic part contains the kinetic energy, the external trapping
potential $U_{\B}(\xV)$ experienced by the bosons, and the $s$-wave
interaction between bosons
\eqmulti{
  \HO_{\B} 
  =& \sum_{i=1}^N \bigg[ \frac{\pOV_i^2}{2 \mB} 
    + U_{\B}(\xOV_i) \bigg]\; \PiO^{\B}_i \\
  &+ \frac{4\pi\,a_{\B0}}{m_{\B}} \sum_{i<j=1}^N \delta^{(3)}(\rOV_{ij})\; 
    \PiO^{\BB}_{ij} ,
}
where $\mB$ is the mass of the bosonic atoms and $a_{\B0}$ the $s$-wave
scattering length of the boson-boson interaction ($\hbar=1$ throughout
this paper). The formal distinction between bosonic and fermionic
atoms is accomplished by projection operators $\PiO^{\B}_i$ and
$\PiO^{\F}_i$ onto bosonic and fermionic single-particle
states. Accordingly, the two-body projection operators
$\PiO^{\BB}_{ij}$, $\PiO^{\FF}_{ij}$, and $\PiO^{\BF}_{ij}$ select
pairs of two bosons, pairs of fermions, and boson-fermion pairs,
respectively.

The structure of the fermionic part of the Hamiltonian differs from
that of the bosonic part: Due to the Pauli principle $s$-wave contact
interactions between identical fermions do not contribute. The lowest
order interaction contribution is therefore the $p$-wave contact
interaction, which will be included here. The operator of the $p$-wave
contact interaction is necessarily nonlocal, i.e., it involves the
relative momentum operator $\qOV_{ij} = \frac{1}{2}(\pOV_i-\pOV_j)$ of
the interacting particle pair. Thus the fermionic Hamiltonian reads
\eqmulti{
  \HO_{\F} 
  =& \sum_{i=1}^N \bigg[ \frac{\pOV_i^2}{2 \mF} 
    + U_{\F}(\xOV_i) \bigg]\; \PiO^{\F}_i \\
  &+ \frac{4\pi\,a_{\F1}^3}{m_{\F}} \sum_{i<j=1}^N 
    \qOV_{ij}\delta^{(3)}(\rOV_{ij})\qOV_{ij}\;\PiO^{\FF}_{ij} ,
}
where $\mF$ is the mass of the fermionic atoms, $U_{\F}(\xV)$ the
respective trapping potential, and $a_{\F1}$ the $p$-wave scattering
length \cite{RoFe01a}. 

Finally, for the interaction between bosons and fermions, both, $s$-wave and
$p$-wave terms are present
\eqmulti{
  \HO_{\BF} 
  =& \frac{4\pi\,a_{\BF0}}{m_{\BF}} \sum_{i<j=1}^N \delta^{(3)}(\rOV_{ij})\; 
     \PiO^{\BF}_{ij} \\
  &+ \frac{4\pi\,a_{\BF1}^3}{m_{\BF}} \sum_{i<j=1}^N 
    \qOV_{ij}\delta^{(3)}(\rOV_{ij})\,\qOV_{ij}\; 
     \PiO^{\BF}_{ij} ,
}
where $m_{\BF} = 2 m_{\B} m_{\F}/(m_{\B}+m_{\F})$ is twice the reduced
mass of a boson-fermion pair, $a_{\BF0}$ is the $s$-wave, and $a_{\BF1}$
the $p$-wave scattering length of the boson-fermion interaction.

In many cases the $p$-wave components of the interactions have only a
marginal effect because the corresponding $p$-wave scattering lengths are
small. Nevertheless, one can expect a significant influence of $p$-wave
interactions on the structure and stability of the mixture provided the
value of the $p$-wave scattering lengths is in the order of the $s$-wave
scattering lengths or larger, e.g. in the vicinity of a $p$-wave Feshbach
resonance \cite{Bohn00}. The strong influence of $p$-wave interactions
on the properties of purely fermionic systems was discussed in detail in
Refs. \cite{RoFe01c,RoFe01a}.

\subsection{Energy functional}

Using the effective Hamiltonian \eqref{eq:mod_hamiltonian} we study
the ground state properties of boson-fermion mixtures at zero
temperature. I adopt the language of density functional theory and
construct, in a first step, an energy functional which connects the
energy expectation value of the mixture with the one-body density
distributions $n_{\B}(\xV)$ and $n_{\F}(\xV)$ of the bosons and the
fermions, respectively,
\eq{
  \matrixe{\Psi}{\HO}{\Psi}  
  \;\to\; 
  E[\nB,\nF] 
  = \int\!\!\dd^3x\;\EC(\xV) .
}
For the construction of the energy functional the different parts of
the Hamiltonian \eqref{eq:mod_hamiltonian} are considered separately
and the energy density $\EC(\xV)$ is decomposed into a purely bosonic
part $\EC_{\B}$, a purely fermionic part $\EC_{\F}$, and a part
$\EC_{\BF}$ that contains the interaction between bosons and fermions
\eq{
  \EC(\xV) 
  =  \EC_{\B}(\xV) + \EC_{\F}(\xV) + \EC_{\BF}(\xV) .
}

Using the many-body state \eqref{eq:mod_state_bosons} the calculation of
the expectation value of the bosonic part of the Hamiltonian $\HO_{\B}$
immediately leads to the well known Gross-Pitaevskii energy density
\cite{DaGi99}
\eqmulti{ \label{eq:mod_energydens_b}
  \EC_{\B}(\xV) 
  =& \frac{1}{2m_{\B}} \big|\nablaV \sqrt{\nB(\xV)}\big|^2
  + U_{\B}(\xV)\,\nB(\xV) \\
  &+ \frac{2\pi\, a_{\B0}}{m_{\B}}\, \nB^2(\xV).
}
The ground state density distribution $\nB(\xV)$ of the bosons is
connected to the single-particle wave function \braket{\xV}{\phi_{\B}}
in \eqref{eq:mod_state_bosons} by
\eq{
  \nB(\xV) 
  = \Phi_{\B}^2(\xV) = N_{\B}\,\braket{\xV}{\phi_{\B}}^2 
}
which reflects the simple structure of the bosonic part of the
many-body state.

For the fermionic part the construction of the associated energy
density is not as straight forward. A general treatment of the
fermionic many-body problem in mean-field approximation requires the
solution of Hartree-Fock equations to obtain the single-particle wave
functions \cite{NyMo99}. For the experimentally realized particle
numbers of $N_{\F}\approx10^6$ a self-consistent numerical treatment
of these differential equations is not feasible.  However, the large
particle numbers allow the use of the Thomas-Fermi approximation for
the fermionic component \cite{RoFe01a}. It is the lowest order of a
semiclassical expansion of the energy density in terms of derivatives
of the single-particle density $n_{\F}(\xV)$ \cite{RiSc80}. All terms
in the energy density that involve gradients or higher order
derivatives of the density are neglected. To obtain the energy density
in Thomas-Fermi approximation one calculates the energy expectation
value of the infinite homogeneous system using a Slater determinant of
all momentum eigenstates up to the Fermi momentum $k_{\F}$ with
periodic boundary conditions. Next, one replaces the constant density
$n_{\F} = k_{\F}^3/(6\pi^2)$ therein by a density distribution
$n_{\F}(\xV)$. This leads to the fermionic contribution to the
energy density \cite{RoFe01a,RoFe01c}
\eqmulti{ \label{eq:mod_energydens_f}
  \EC_{\F}(\xV)
  =& \frac{3(6\pi^2)^{2/3}}{10 m_{\F}}\, \nF^{5/3}(\xV)
  + U_{\F}(\xV)\,\nF(\xV) \\
  &+ \frac{(6\pi^2)^{5/3}\, a_{\F1}^3}{5\pi m_{\F}}\, \nF^{8/3}(\xV) ,
}  
where the first term represents the kinetic energy, the second term the
external trapping potential, and the third term the $p$-wave
fermion-fermion interaction.

One can employ the Thomas-Fermi approximation also to calculate the
contribution of the boson-fermion interaction to the energy density
\eqmulti{ \label{eq:mod_energydens_bf}
  \EC_{\BF}(\xV)
  =& \frac{4\pi\,a_{\BF0}}{m_{\BF}}\, \nB(\xV)\nF(\xV) \\
  &+ \frac{(6\pi^2)^{5/3} a_{\BF1}^3}{10\pi m_{\BF}}\, \nB(\xV) \nF^{5/3}(\xV) ,
}
where the first term describes the $s$-wave and the second term the
$p$-wave boson-fermion interaction.

Notice that the Thomas-Fermi approximation affects only the contributions
to the energy density associated with nonlocal terms of the Hamiltonian,
that is the kinetic energy and the $p$-wave interactions. For fermion
numbers in the order $N_{\F}\approx1000$ one can check explicitly that the
Thomas-Fermi approximation is in good agreement with Hartree-Fock type
calculations \cite{NyMo99}. Alternatively one can evaluate the next terms
of the semiclassical gradient expansion of the energy density
\cite{RoFe01a}. Both checks show that the Thomas-Fermi approximation
generally yields a very good description of the fermionic component for
$N_{\F}\gtrsim1000$.

Some authors \cite{Molm98,AkVi02} investigated the structure of
boson-fermion mixtures using the Thomas-Fermi approximation also for the
bosonic species. Here, however, one encounters the problem that the
kinetic energy contribution of the bosons vanishes completely (not so for
the kinetic energy of the fermions). This causes two stringent conditions
for the applicability of the approximation: (a) the boson-boson
interaction has to be repulsive, i.e., systems with vanishing or
attractive boson-boson interaction cannot be considered at all. (b) The
contribution of the boson-boson interaction to the energy density has to
be much larger than the neglected kinetic energy. For the investigation of
the phase diagram as presented in the following sections this means a
substantial and uncontrolled limitation of the parameter range in which
the approximation may be valid \cite{RoFe02}. Therefore, I employ the
Thomas-Fermi approximation only for the fermionic component, where it is
well suited for all types of interactions, and treat the bosons using the
full Gross-Pitaevskii energy density.

\subsection{Euler-Lagrange equations}

The second step is the minimization of the total energy
$E[n_{\B},n_{\F}]$ by functional variation of the boson and fermion
density distribution. The variation shall be performed under the
constraint of given numbers of bosons and fermions, $N_{\B}$ and
$N_{\F}$. The constraints are implemented by two Lagrange multipliers,
i.e. the chemical potentials $\mu_{\B}$ and $\mu_{\F}$, and an
unrestricted variation is carried out for the Legendre transformed
energy
\eqmulti{ \label{eq:mod_energyfunc_trans}
  F[n_{\B},n_{\F}] 
  &= E[n_{\B},n_{\F}] - \mu_{\B} N[\nB] - \mu_{\F} N[\nF] \\
  &= \int\!\!\dd^3x\; \FC(\xV)
}
with a transformed energy density
\eq{
  \FC(\xV)
  = \EC(\xV) - \mu_{\B} \nB(\xV) - \mu_{\F} \nF(\xV) .
}
The values of the chemical potentials are chosen (a posteriori) such
that the integrals over the boson ($\xi=\B$) and the fermion densities
($\xi=\F$)
\eq{
  N[n_{\xi}] 
  = \int\!\!\dd^3x\; n_{\xi}(\xV)
}
yield the desired particle numbers $N_{\B}$ and $N_{\F}$,
respectively.

A necessary condition for the density distributions
$\nB(\xV)=\PhiB^2(\xV)$ and $\nF(\xV)$ to describe a minimum of the
transformed energy $F[n_{\B},n_{\F}]$ is given by the Euler-Lagrange
equations
\eq{
  0 = \frac{\partial\FC}{\partial\PhiB} 
    - \nablaV \frac{\partial\FC}{\partial(\nablaV\PhiB)}
  \;,\quad
  0 = \frac{\partial\FC}{\partial\nF} 
    - \nablaV \frac{\partial\FC}{\partial(\nablaV\nF)} .
}
One can insert the explicit form of $\FC(\xV)$, which involves the
energy densities \eqref{eq:mod_energydens_b},
\eqref{eq:mod_energydens_f}, and \eqref{eq:mod_energydens_bf}, and
obtain two coupled equations that determine the ground state densities
$\nB(\xV)$ and $\nF(\xV)$.

The Euler-Lagrange equation for the boson density reduces to a
modified Gross-Pitaevskii equation
\eqmulti{ \label{eq:mod_eulerlag_boson}
 0 
 = \bigg[&-\frac{1}{2\mB} \nablaV^2 + U_{\B}(\xV) - \mu_{\B} 
   + \frac{4\pi\,a_{\B0}}{\mB}\,\PhiB^2(\xV) \\
 &+ \frac{4\pi\,a_{\BF0}}{\mBF}\,\nF(\xV) 
   + \frac{(6\pi^2)^{5/3} a_{\BF1}^3}{10\pi m_{\BF}}\, \nF^{5/3}(\xV)
   \bigg] \Phi_{\B}(\xV) .
}
In addition to the usual terms describing kinetic energy, trapping
potential, chemical potential, and boson-boson interaction one obtains
two contributions from the $s$- and $p$-wave interaction between bosons
and fermions. The bosons thus experience additional mean-field
potentials generated by the boson-fermion interactions that depend on
different powers of the fermionic density distribution.

For the fermions the Euler-Lagrange equation simplifies to a
polynomial equation in $\nF(\xV)$ because all terms in the energy
density that involve the fermion density are local --- a direct
consequence of the Thomas-Fermi approximation
\eqmulti{ \label{eq:mod_eulerlag_fermion}
 0 
 =& \bigg[ U_{\F}(\xV) - \mu_{\F} 
   + \frac{4\pi a_{\BF0}}{\mBF}\,\nB(\xV) \bigg] \\
 &+ \bigg[ \frac{(6\pi^2)^{2/3}}{2 \mF} 
   + \frac{(6\pi^2)^{5/3} a_{\BF1}^3}{6\pi \mBF}\, \nB(\xV)\bigg]
   \nF^{2/3}(\xV) \\
 &+ \bigg[ \frac{8(6\pi^2)^{5/3} a_{\F1}^3}{15\pi \mF} \bigg]
   \nF^{5/3}(\xV) .
}
Again the boson density enters explicitly through the $s$- and $p$-wave
boson-fermion interaction. The simultaneous solution of
Eqs. \eqref{eq:mod_eulerlag_boson} and \eqref{eq:mod_eulerlag_fermion}
yields density profiles that describe a stationary point of the energy
functional $F[\nB,\nF]$. In principle one has to check explicitly that
these coincide with a minimum of the energy.

\subsection{Numerical treatment}

On the basis of the coupled equations \eqref{eq:mod_eulerlag_boson}
and \eqref{eq:mod_eulerlag_fermion} the structure of binary
boson-fermion mixtures is studied with special emphasis on the
effects and the interplay between the boson-boson and the
boson-fermion $s$-wave interaction.

In order to reduce the number of physical parameters it is assumed
that the $p$-wave scattering lengths are small and the corresponding
$p$-wave terms are neglected. Thus the Euler-Lagrange equations for the
boson and fermion density simplify to
\eqmulti{ \label{eq:mod_gp_boson}
  0 
  = \bigg[&-\frac{1}{2\mB} \nablaV^2 + U_{\B}(\xV) - \mu_{\B} 
    + \frac{4\pi\,a_{\B}}{\mB}\,\PhiB^2(\xV) \\
  &+ \frac{4\pi\,a_{\BF}}{\mBF}\,\nF(\xV) \bigg] \Phi_{\B}(\xV) 
}
for the boson density $\nB(\xV) = \PhiB^2(\xV)$ and
\eq{ \label{eq:mod_tf_fermion}
  \nF(\xV) 
  = \frac{(2\mF)^{3/2}}{6\pi^2}
    \bigg[\mu_{\F} - U_{\F}(\xV) - \frac{4\pi\, a_{\BF}}{\mBF} \nB(\xV) 
    \bigg]^{3/2}
}
for the fermion density. The index ``0'' which marked the $s$-wave
scattering lengths is omitted here and in the following.
Furthermore, we will restrict ourselves to parabolic trapping
potentials with spherical symmetry
\eq{
  U_{\xi}(\xV)
  = \frac{m_{\xi} \omega_{\xi}^2}{2} x^2
  = \frac{1}{2 m_{\xi} \ell_{\xi}^4} x^2 ,
}
where $\omega_{\xi}$ is the oscillator frequency and
$\ell_{\xi}=(m_{\xi}\omega_{\xi})^{-1/2}$ is the corresponding oscillator
length for the bosonic ($\xi=\B$) and fermionic component ($\xi=\F$),
respectively. In a magnetic trap the trap parameters
of the fermionic species can be obtained from those for the bosons by
scaling according to the different masses and magnetic moments.

For the numerical treatment it is convenient to express all quantities
with the dimension of length in units of the oscillator length
$\ell_{B}$ and energies in units of the oscillator frequency
$\omega_{\B}$. Eventually, the physically relevant dimensionless
parameters to characterize the mixture are the particle numbers
$N_{\B}$ and $N_{\F}$, the ratios of the scattering lengths and the
oscillator length $a_{\B}/\ell_{\B}$ and $a_{\BF}/\ell_{\B}$, and the
ratios of the masses $m_{\B}/m_{\F}$ and of the oscillator lengths
$\ell_{\B}/\ell_{\F}$.

The solution of the coupled equations \eqref{eq:mod_gp_boson} and
\eqref{eq:mod_tf_fermion} is accomplished by an efficient imaginary
time propagation algorithm --- also known as quantum diffusion
algorithm \cite{Feag94} --- which is embedded in a simple iterative
scheme. A basic iteration step consists of two stages: (a) The
fermionic density profile $n_{\F}(\xV)$ is calculated from
\eqref{eq:mod_tf_fermion} with a chemical potential $\mu_{\F}$
adjusted such that the desired fermion number $N_{\F}$ is reproduced.
(b) A single imaginary time step is performed using the mean-fields
associated with the fermion density determined before and the boson
density of the previous iteration step. The resulting $n_{\B}(\xV)$ is
then normalized to the total boson number $N_{\B}$ and used as
initialization for the next iteration step.  In the beginning of the
iteration cycle the boson density is initialized with a Gaussian
profile chosen, e.g., according to a variational treatment of the
isolated bosonic system.

Typically $200$ grid points are used to set up a radial lattice in the
region where the bosonic and the fermionic density distributions
overlap. The fermionic density profile outside the overlap region
(where $\nB(\xV)\equiv 0$) is described using the analytic form
\eqref{eq:mod_tf_fermion}. The imaginary time propagation itself is
implemented using either a linearized time-evolution operator with a
simple three-point discretization of the radial part of the Laplacian
or by a split operator technique employing a fast Fourier
transformation. The size of a time step is chosen sufficiently small
to guarantee the stability of the algorithm and the independence of
the results. The convergence of the density profiles is monitored
during the evolution.

\section{Attractive boson-fermion interactions}
\label{sec:BF-}

In this section generic properties of binary boson-fermion mixture
with attractive interactions between the two species ($a_{\BF}<0$) are
discussed. Recent experiments indicate that mixtures of bosonic
${}^{87}\text{Rb}$ and fermionic ${}^{40}\text{K}$ atoms belong to
this class of interactions \cite{RoRi02,FeIn02,GoPa02}.

Throughout this and the following section it is assumed that the
masses ($m=m_{\B}=m_{\F}$) and the oscillator lengths of the trapping
potentials ($\ell = \ell_{\B} = \ell_{\F}$) are identical for both
components in order to reduce the parameter manifold to a tractable
size.

\subsection{Density profiles}
\label{sec:BF-_density}

\begin{figure}
\includegraphics[height=0.23\textheight]{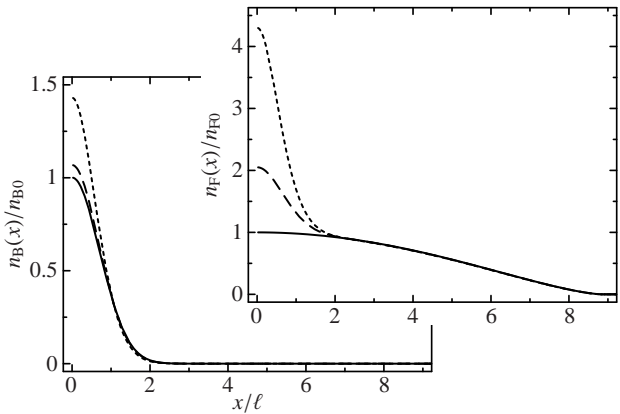}
\caption{Boson (lower panel) and fermion density profiles (upper
  panel) for a mixture with $N_{\B}=N_{\F}=10^4$ and $\aB/\ell=0$ for
  different values of the boson-fermion scattering length:
  $\aBF/\ell=0.0$ (solid lines), $-0.001$ (dashed), and $-0.002$
  (dotted). Both densities are given in units of the respective
  central densities $n_{\B0}=N_{\B}/(\pi^{3/2}\ell^3)\approx 1796\, 
  \ell^{-3}$ and $n_{\F0}=(48 N_{\F})^{1/2}/(6\pi^2\ell^3) \approx
  11.7\,\ell^{-3}$ of the noninteracting system with the same 
  particle numbers.}
\label{fig:BF-_densities_varBF_1}
\end{figure}

Some of the basic properties of mixtures with attractive boson-fermion
interactions become evident from the shape of the density profiles
already.  Figure \ref{fig:BF-_densities_varBF_1} shows the radial
density profiles for mixtures with $N_{\B}=N_{\F}=10^4$, $\aB/\ell=0$,
and different strengths of the attractive boson-fermion
interaction. Already the noninteracting case $\aBF/\ell=0$ (solid
tlines) reveals a fundamental property of these mixtures: The fermionic
density distribution has a much larger spatial extension than a
bosonic distribution with the same particle number.  This is a direct
consequence of the Pauli principle that forces the fermions to occupy
``excited'' single-particle states of the external potential which
have a larger radial extension whereas the bosons in a Bose-Einstein
condensate all occupy the ground state of the trap. This manifestation
of the so-called Fermi pressure was observed experimentally
\cite{TrSt01}.

The inclusion of an attractive boson-fermion interaction leads to an
enhancement of the boson and the fermion density within the central
overlap region as it is shown by the broken curves in
Fig. \ref{fig:BF-_densities_varBF_1}. The bosonic profile narrows and
the central density increases moderately. The effect on the fermionic
component is more pronounced: within the overlap region the fermion
density exhibits a high-density bump on top of the low-density
profile. The central fermion density can easily be increased by a
factor of four (dotted line in Fig. \ref{fig:BF-_densities_varBF_1}).
This structure can be understood by considering the mean-field
potential experienced by the fermions: In addition to a shallow
trapping potential the boson-fermion attraction generates a tight
potential well with the shape of the bosonic density distribution
which causes a localized enhancement of the fermion density profile.
\begin{figure}
\includegraphics[height=0.23\textheight]{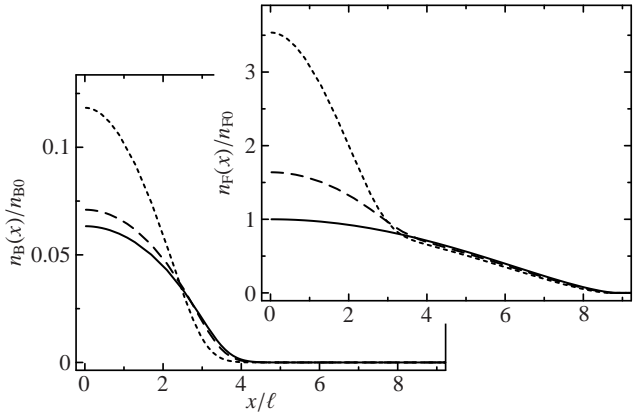}
\caption{Boson and fermion density profiles for a mixture with
  $N_{\B}=N_{\F}=10^4$ and $\aB/\ell=0.005$ for different values of
  the boson-fermion scattering length: $\aBF/\ell=0.0$ (solid lines),
  $-0.01$ (dashed), and $-0.02$ (dotted). The densities are given in units
  of the central boson and fermion density of the noninteracting system 
  (compare Fig. \ref{fig:BF-_densities_varBF_1}). }
\label{fig:BF-_densities_varBF_2}
\end{figure}

The boson-boson interaction has a strong influence on this density
enhancement. The presence of a boson-boson repulsion broadens the
bosonic distribution and reduces the maximum density
significantly. This in turn strongly reduces the enhancement of the
fermion density in the overlap region. As is shown in
Fig. \ref{fig:BF-_densities_varBF_2} much stronger boson-fermion
attractions are required to generate a similar increase of the
fermion density in the presence of repulsive boson-boson interactions.
However, the overlap volume is much larger in these cases; thus a
larger fraction of the fermions is contained in the high-density
region.

\subsection{Simultaneous mean-field collapse}
\label{sec:BF-_collapse}

For given particle numbers $N_{\B}$ and $N_{\F}$ and given
boson-boson scattering length $a_{\B}$ there exists a maximum strength
of the attractive boson-fermion interaction up to which the mixture is
stable. Physically this stability limit is governed by the balance
between the kinetic energy of bosons and fermions and the mutually
attractive mean-field generated by the boson-fermion interaction.  If
the boson-fermion attraction becomes too strong --- likewise if the
boson or fermion numbers become too large --- the attractive mean-field
cannot be stabilized by the kinetic energy anymore and the mixture can
lower its energy by increasing the boson and fermion density; both
density distributions collapse simultaneously within the overlap
region. 

In the numerical treatment of the coupled Gross-Pitaevskii problem the
instability is indicated by a divergence of the central boson and
fermion density during the imaginary time propagation. By monitoring
the central density one can determine whether the systems is stable or
collapses, i.e., whether the density converges or diverges for
a given set of parameters $N_{\B}$, $N_{\F}$, $a_{\B}/\ell$, and
$a_{\BF}/\ell$. To determine the stability limit in presence of
attractive boson-fermion interactions it is most useful to vary
$a_{\BF}/\ell$ for fixed values of $N_{\B}$, $N_{\F}$, and $a_{\B}/\ell$
until the onset of instability is reached.

\begin{figure}
\includegraphics[height=0.58\textheight]{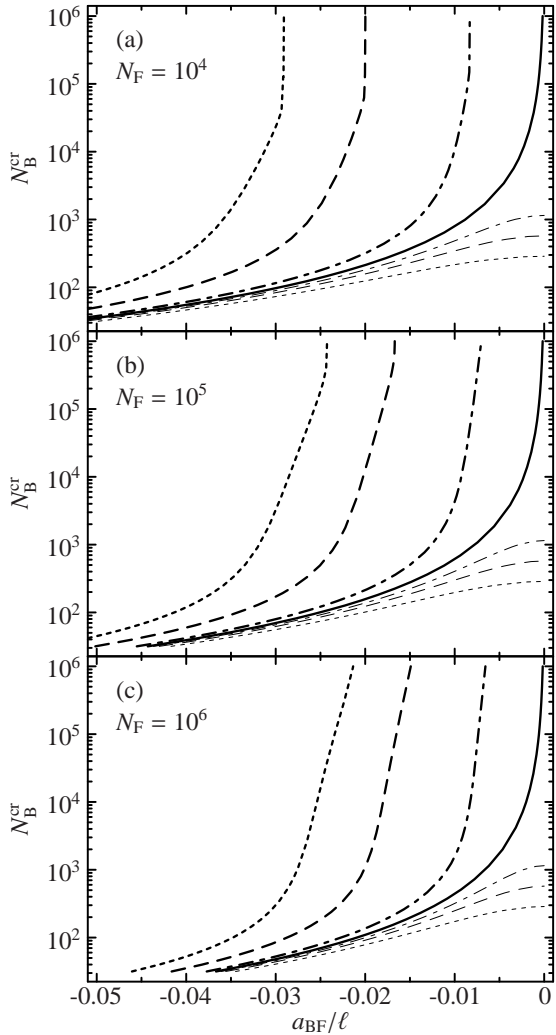}
\caption{Critical number of bosons $N_{\B}^{\crit}$
  above which the mixture collapses simultaneously as function of
  $a_{\BF}/\ell<0$. The thick curves correspond to repulsive
  boson-boson interactions: $a_{\B}/\ell=0$ (full line), $0.001$
  (dash-dotted), $0.005$ (dashed), $0.01$ (dotted). Thin lines show
  attractive boson-boson interactions: $a_{\B}/\ell = -0.0005$ (thin
  dash-dotted), $-0.001$ (thin dashed), $-0.002$ (thin dotted). The
  three panels represent different fermion numbers: (a) $N_{\F}=10^4$,
  (b) $N_{\F}=10^5$, (c) $N_{\F}=10^6$. }
\label{fig:BF-_collapse_NNBcrit}
\end{figure}

Figure \ref{fig:BF-_collapse_NNBcrit} depicts the resulting stability
limit in terms of a critical number of bosons $N_{\B}^{\crit}$ as
function of $a_{\BF}/\ell<0$ for different $a_{\B}/\ell$ and
$N_{\F}$. To obtain the curves I determined the limiting value of
$a_{\BF}/\ell$ that still allows a stable mixture for a set of boson
numbers $\log_{10} N_B=1.5,...,6$. The maximum strength of the
boson-fermion interaction of $a_{\BF}/\ell=-0.05$ shown in the plots
corresponds to a scattering length of $a_{\BF}\approx
-1000\,a_{\text{Bohr}}\approx -50\,\text{nm}$ for a typical trap with
$\ell=1\,\mu\text{m}$.

The full curves in Fig. \ref{fig:BF-_collapse_NNBcrit} show the
critical boson number for a mixture with vanishing boson-boson
interaction ($a_{\B}/\ell=0$). Already a moderate boson-fermion
attraction with $a_{\BF}/\ell=-0.005$ causes a severe limitation of
the boson number to $N_{B}\lesssim2300$ in the presence of
$N_{\F}=10^5$ fermions, as it is shown in panel (b) of
Fig. \ref{fig:BF-_collapse_NNBcrit}.

The boson-boson interaction has a very strong influence on the
structure and stability of the mixture. If a weak boson-boson
repulsion with a fraction of the strength of the boson-fermion
attraction is included the system is stabilized significantly,
i.e., the critical number of boson $N_{\B}^{\crit}$ is increased. This
is depicted by the broken curves in
Fig. \ref{fig:BF-_collapse_NNBcrit} which correspond to different
strengths $a_{\B}/\ell>0$ of the repulsive boson-boson interaction.
For the previous example ($a_{\BF}/\ell=-0.005$, $N_{\F}=10^5$) a
boson-boson repulsion with $a_{\B}/\ell=0.001$ is sufficient to
increase the maximum number of bosons in a stable system from
$N_{\B}^{\crit}\approx2300$ for $a_{\B}/\ell=0$ to far more than
$10^6$. Moreover, repulsive boson-boson interactions can cause the
absolute stabilization of the mixture against collapse induced by the
boson-fermion attraction. That is, for given $a_{\B}/\ell>0$ and
$N_{\F}$ mixtures with $|a_{\BF}/\ell|$ below a certain limiting
strength are stable up to any number of bosons. This can be clearly
seen in Fig. \ref{fig:BF-_collapse_NNBcrit}(a): the broken thick
curves exhibit a kink and start to rise vertically at a specific value
of $a_{\BF}/\ell$ which depends on $a_B/\ell$ and $N_F$. All systems
with weaker boson-fermion attraction are stable against collapse for
arbitrary $N_{\B}$.

An attractive boson-boson interaction, on the other hand, promotes the
collapse of the mixture. It generates an additional attractive
contribution to the mean-field experienced by the bosons which seeks
to increase the boson density and has to be balanced by the kinetic
energy. Due to the boson-fermion attraction the fermionic component in
turn feels a stronger mean-field attraction cause by the increased
boson density. Therefore the critical boson number $N_{\B}^{\crit}$ is
lowered by attractive boson-boson interactions as the thin curves in
Fig. \ref{fig:BF-_collapse_NNBcrit} show. One has to be aware that the
character of the collapse changes if one reduces $a_{\BF}/\ell\to0$
for fixed $a_{\B}/\ell<0$. For reasonably strong boson-fermion
attraction the mutual mean-field couples the boson and fermion density
strongly to each other such that both densities collapse simultaneously
in the overlap region. If the boson-fermion interaction is reduced
this coupling weakens until bosons and fermions decouple for
$a_{\BF}/\ell=0$. In this case the boson density may still collapse
due to the boson-boson attraction; the fermion density, however, is
not affected and remains stable (see also Sec. \ref{sec:BF+_collapse}).

The number of fermions in the system has a rather weak influence on
the stability. The three panels in Fig. \ref{fig:BF-_collapse_NNBcrit}
correspond to (a) $N_{\F}=10^4$, (b) $N_{\F}=10^5$, and (c)
$N_{\F}=10^6$. Mainly, the increasing number of fermions $N_{\F}$
enhances the effect of the boson-fermion attraction. Thus the
stability limit for a given $N_{\B}$ is shifted towards smaller values
of $a_{\BF}/\ell$ if $N_{\F}$ is increased.

\section{Repulsive boson-fermion interactions}
\label{sec:BF+}

In this section the generic properties of boson-fermion mixtures with
repulsive interactions between the two species ($a_{\BF}>0$) are
discussed. All ${}^7\text{Li}$-${}^6\text{Li}$ mixtures used in
present experiments \cite{TrSt01, ScFe01, ScKh01} belong to this class
of interactions.

\subsection{Density profiles}
\label{sec:BF+_density}

\begin{figure}
\includegraphics[height=0.23\textheight]{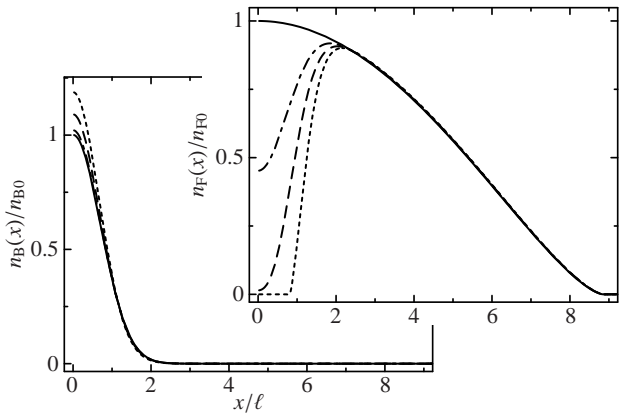}
\caption{Boson (lower panel) and fermion density profiles (upper
  panel) for a mixture with $N_{\B}=N_{\F}=10^4$ and $\aB/\ell=0$
  for different values of the boson-fermion scattering length:
  $\aBF/\ell=0.0$ (solid lines), $0.0007$ (dash-dotted), $0.0015$
  (dashed), and $0.003$ (dotted). Both densities are given in units 
  of the central densities of the corresponding noninteracting system as in
  Fig. \ref{fig:BF-_densities_varBF_1}.}
\label{fig:BF+_densities_varBF_1}
\end{figure}

Obviously a repulsive boson-fermion interaction will tend to reduce
the overlap between the two species --- in contrary to the attractive
boson-fermion interactions discussed in the previous section. Hence
the structural transition characteristic for this class of
interactions is the spatial separation (or demixing) of the two
species \cite{NyMo99,Molm98}. Figure \ref{fig:BF+_densities_varBF_1}
shows an example for the influence of repulsive boson-fermion
interactions of increasing strength on the density profiles of the
mixture. The compact boson distribution in the trap center repels the
fermionic cloud from the overlap region. With increasing boson-fermion
scattering length $a_{\BF}/\ell>0$ the fermionic distribution gets
more and more depleted until $n_{\F}(\xV=0)$ reaches zero (see the
dashed curve in Fig. \ref{fig:BF+_densities_varBF_1}). For even larger
$a_{\BF}/\ell$ the fermions are expelled farther from the trap center
and the overlap of both species is reduced continuously. The boson
cloud in return is slightly compressed by the outer shell of fermions.
\begin{figure}
\includegraphics[height=0.23\textheight]{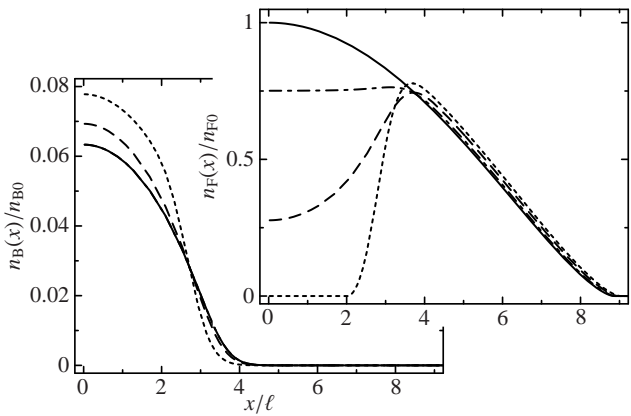}
\caption{Boson (lower panel) and fermion density profiles (upper
  panel) for a mixture with $N_{\B}=N_{\F}=10^4$ and $\aB/\ell=0.005$ for
  different values of the boson-fermion scattering length:
  $\aBF/\ell=0.0$ (solid lines), $0.005$ (dash-dotted), $0.015$
  (dashed), and $0.03$ (dotted). The densities are given in units of the
  central densities $n_{\B0}$ and $n_{\F0}$, resp., of the noninteracting
  system with the same particle numbers (compare to Fig. 
  \ref{fig:BF-_densities_varBF_1}). The dash-dotted curve for the boson
  density is on top of the solid curve.}
\label{fig:BF+_densities_varBF_2}
\end{figure}

The boson-boson interaction has a very strong influence on the density
profiles of bosons and fermions. Similar to the case discussed in
Sec. \ref{sec:BF-_density} a repulsive boson-boson interaction
reduces the influence of the boson-fermion interaction. Much larger
boson-fermion scattering lengths are required to cause phase
separation in the presence of repulsive boson-boson interactions. An
example with $a_{\B}/\ell=0.005$ is shown in Fig.
\ref{fig:BF+_densities_varBF_2}. Because the bosonic distribution is
broadened due to the boson-boson repulsion, the fermion density is
depleted in a significantly larger volume. A peculiar structure
appears for interactions with $a_{\B}/\ell = a_{\BF}/\ell >0$: The
fermion density is nearly constant within the whole overlap region as
the dash-dotted curve in Fig.  \ref{fig:BF+_densities_varBF_2}
shows \cite{Molm98}. 

Depending on the trap geometry the separated phase can exhibit
different structures. If the fermionic species experiences a tighter
confinement than the bosonic component (requires an appropriate
combination of magnetic moments) an inversion of the separated
configuration can appear: a central core of fermions surrounded by a
thin boson shell which compresses the fermions. In these cases there
exists a second structural transition from the fermion-core
configuration at smaller $N_{\B}$ to the usual boson-core
configuration at large $N_{\B}$.  In deformed traps a rich variety of
asymmetric configurations are possible, some of these were discussed
in \cite{NyMo99}.

\subsection{Spatial component separation}
\label{sec:BF+_separation}

In this section the influence of scattering lengths and particle
numbers on the transition from overlapping to separated density
distributions is quantified. Since the transition is smooth in most
cases there is no unique definition for the onset of separation. As a
gross criterion for separation I use the fermion density in the trap
center and consider those configurations as separated that yield
$n_{\F}(\xV=0)=0$. Clearly, if the boson-fermion repulsion is
increased then the overlap of the two density distributions decreases
further.

\begin{figure}
\includegraphics[height=0.58\textheight]{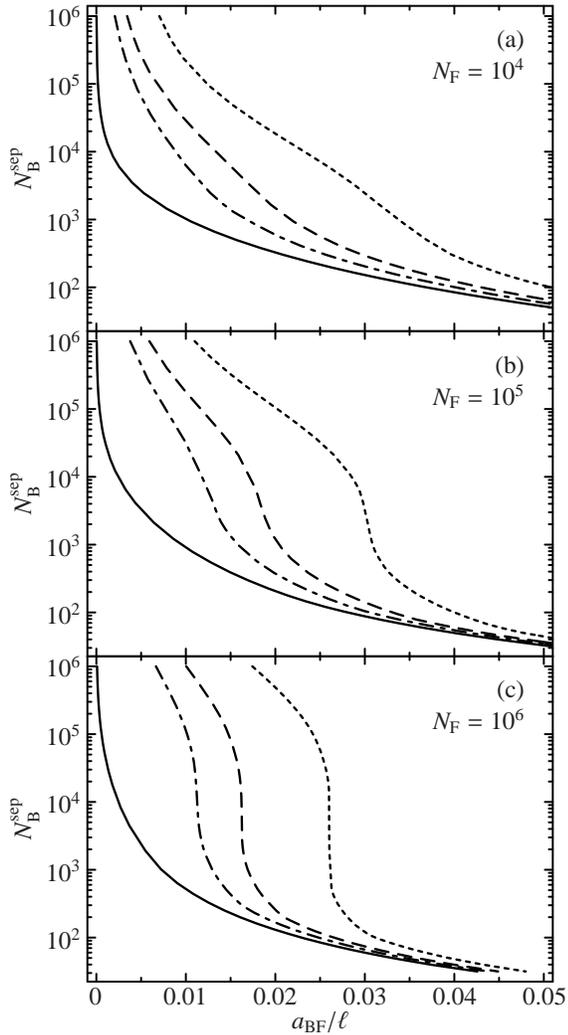}
\caption{Characteristic boson number $N_{\B}^{\sep}$
  for component separation as function of $a_{\BF}/\ell>0$. The curves
  correspond to different values of the boson-boson scattering
  length: $a_{\B}/\ell=0$ (full line), $0.001$ (dash-dotted), $0.002$
  (dashed), $0.005$ (dotted). The three panels represent different
  fermion numbers: (a) $N_{\F}=10^4$, (b) $N_{\F}=10^5$, (c)
  $N_{\F}=10^6$.  }
\label{fig:BF+_separation_NNBsep}
\end{figure}

Figure \ref{fig:BF+_separation_NNBsep} summarizes the findings on the
characteristic number of bosons $N_{\B}^{\sep}$ above which the
fermion density vanishes in the trap center. The curves correspond to
different values of the boson-boson scattering length $a_{\B}/\ell$.
For increasing boson-fermion repulsion $a_{\BF}/\ell>0$ the
characteristic boson number $N_{\B}^{\sep}$ drops rapidly. In the case
$a_{\B}/\ell=0$ (solid lines) and $N_{\F}=10^5$ a moderate boson-fermion
repulsion with $a_{\BF}/\ell=0.005$ leads to phase separation  
if the number of bosons exceeds $N_{\B}^{\sep} \approx 3300$.
 
The inclusion of repulsive boson-boson interactions leads to an
interesting modification of the separation behavior. First of all the
boson-boson repulsion stabilizes the mixture against phase separation,
i.e., $N_{\B}^{\sep}$ increases significantly with increasing
$a_{\B}/\ell>0$ as the broken curves in
Fig. \ref{fig:BF+_separation_NNBsep} indicate.  Moreover, for
sufficiently large fermion numbers the boson-boson repulsion causes a
qualitative change of the dependence of $N_{\B}^{\sep}$ on the
boson-fermion scattering length. An example is shown in Fig.
\ref{fig:BF+_separation_NNBsep}(c) for $N_{\F}=10^6$. Around a
characteristic value of $a_{\BF}/\ell$, which increases with
increasing $a_{\B}/\ell$, the limiting boson number for separation
$N_{\B}^{\sep}$ suddenly drops by one or more orders of magnitude and
a steplike structure appears.  For values of $a_{\BF}/\ell$ smaller
than the characteristic value associated with this step the
boson-boson repulsion causes a substantial stabilization of the system
against separation. Above the step the influence of the boson-boson
repulsion on $N_{\B}^{\sep}$ is weak.

In addition to the particular behavior of $N_{\B}^{\sep}$ the
character of the transition changes. One may regard the central
fermion density $\nF(0)$ as order parameter; finite values of the
order parameter characterize the overlapping phase, a vanishing order
parameter indicates separation. In most cases, e.g. for
$a_{\B}/\ell=0$, the order parameter decreases continuously with
increasing $N_{\B}$, i.e., the central fermion density decreases
smoothly until the separated phase with $\nF(0)=0$ is reached.  This
behavior corresponds to a second order or cross-over transition.
However, for sufficiently large $N_{\F}$ and $a_{\B}/\ell$, e.g.,
$N_{\F}=10^6$ and $a_{\B}/\ell=0.005$ [see dotted curve in
Fig. \ref{fig:BF+_separation_NNBsep}(c)], the order of the phase
transition changes. For values of $a_{\BF}/\ell$ above the step the
transition is of first order: With increasing $N_{\B}$ the order
parameter decreases slightly until $N_{\B}^{\sep}$ is reached, then it
drops discontinuously to zero. For $a_{\BF}/\ell$ below the value
associated with the step the transition is still continuous which
implies the existence of a tricritical point at the step.

\subsection{Collapse of the bosonic component}
\label{sec:BF+_collapse}

If the repulsive boson-fermion interaction is supplemented by a
boson-boson attraction ($a_{\B}<0$) then a subtle competition between
component separation and mean-field collapse occurs.

For vanishing boson-fermion scattering length $a_{\BF}=0$ the two
species decouple. The bosonic component may undergo a mean-field
collapse just like an isolated Bose-Einstein condensate if the
boson-boson interaction is attractive. The critical particle number
for this collapse of the bosonic component in a decoupled mixture can
be parameterized by $N_{\B,a_{\BF}=0}^{\crit} = 0.575\,
\ell/|a_{\B}|$. This coincides with the result for the collapse a pure
Bose gas \cite{DaGi99,ElHu00} which was basically confirmed by
experiment \cite{RoCl01,BrSa97}.

One should notice that the character of the collapse induced by a
boson-boson attraction differs from that of the collapse cause by an
attractive boson-fermion interaction discussed in
Sec. \ref{sec:BF-_collapse}. In the latter case the boson-fermion
interaction generates a mutual attractive mean-field that acts on
both, fermions and bosons, and causes a simultaneous collapse of both
density distributions. In the case of a boson-boson attraction only
the bosonic component experiences an attractive mean-field and may
become unstable. For $a_{\BF}>0$ the fermionic density profile will be
influenced by a collapse of the bosonic component but it will not
collapse itself. The collapse of the bosonic component may, e.g.,
generate a collective excitation of the fermionic cloud.

\begin{figure}
\includegraphics[height=0.58\textheight]{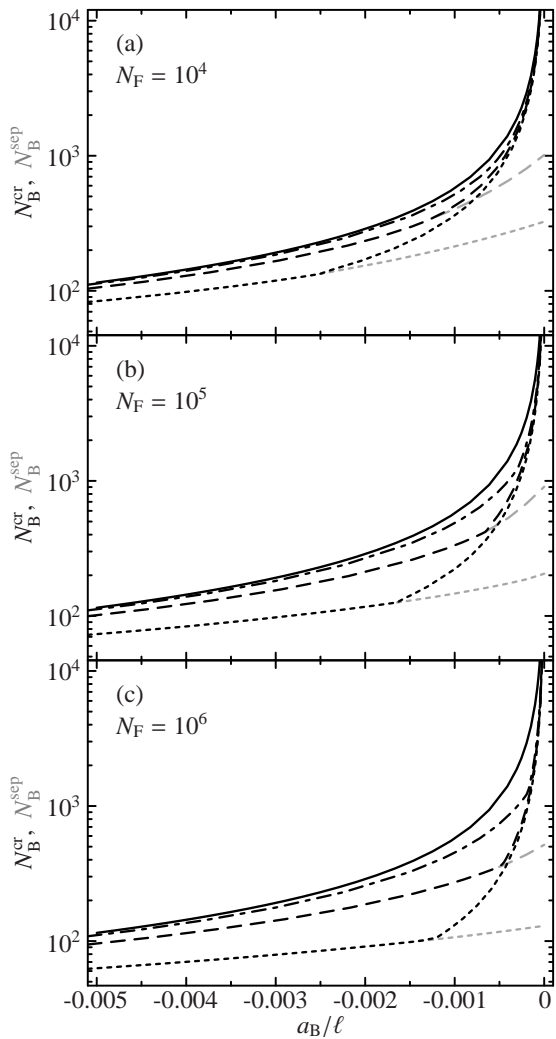}
\caption{Critical number of bosons $N_{\B}^{\crit}$
  above which the bosonic component collapses as function of
  $a_{\B}/\ell<0$ (black lines) as well as the characteristic boson
  number for component separation $N_{\B}^{\sep}$ (gray lines). The
  curves correspond to different values of the boson-fermion
  scattering length: $a_{\BF}/\ell=0$ (full line), $0.005$
  (dash-dotted), $0.01$ (dashed), $0.02$ (dotted). The three panels
  represent different fermion numbers: (a) $N_{\F}=10^4$, (b)
  $N_{\F}=10^5$, (c) $N_{\F}=10^6$.  }
\label{fig:BF+_collapse_NNBcrit}
\end{figure}

Although the collapse for $a_{\B}<0$ concerns the bosonic density
alone, the boson-fermion interaction modifies the stability properties
significantly. We will concentrate on the case $a_{\BF}>0$; the
collapse in presence of attractive boson-fermion interactions was
discussed in Sec. \ref{sec:BF-_collapse} already. The black curves in
Fig. \ref{fig:BF+_collapse_NNBcrit} show the critical number of bosons
$N_{\B}^{\crit}$ as function of the boson-boson scattering length
$a_{\B}/\ell$ for different values of the boson-fermion scattering
length $a_{\BF}/\ell \ge 0$. The solid lines show the critical boson
numbers for the decoupled mixture ($a_{\BF}=0$).

The first general observation is that a boson-fermion repulsion
\emph{de}stabilizes the bosonic component, i.e., the critical boson number
$N_{\B}^{\crit}$ is reduced if $a_{\BF}/\ell$ or $N_{\F}$ are
increased. Physically this can be understood from the structure of the
density profiles discussed in Sec. \ref{sec:BF+_density}. Due to the
boson-fermion repulsion the outer fermionic cloud tends to compress the
compact bosonic core. The boson density and thus the attractive
mean-field are enhanced which eventually promotes the collapse of the
bosonic component. However, as can be seen from
Fig. \ref{fig:BF+_collapse_NNBcrit} the boson-fermion scattering
length $a_{\BF}/\ell$ has to be significantly larger than
$|a_{\B}/\ell|$ to have a noticeable effect.

Evidently, component separation due to a boson-fermion repulsion has a
large effect on the stability of the mixture with respect to collapse
induced by the boson-boson attraction. The gray lines in
Fig. \ref{fig:BF+_collapse_NNBcrit} indicate the characteristic
particle number $N_{\B}^{\sep}$ for the onset of component separation
for interactions with $a_{\BF}/\ell=0.01$ and $0.02$. For weak
attractive boson-boson interactions separation happens at lower
particle numbers than the collapse of the bosonic component, that is
$N_{\B}^{\crit}>N_{\B}^{\sep}$. With increasing strength of the
boson-boson attraction the critical particle number for collapse drops
rapidly until $N_{\B}^{\crit}=N_{\B}^{\sep}$. From that point on
collapse and separation happen simultaneously, i.e. there is no stable
separated configuration anymore. The corresponding curves for
$N_{\B}^{\crit}$ shown in Fig. \ref{fig:BF+_collapse_NNBcrit} exhibit
a kink at this point and proceed towards larger $|a_{\B}/\ell|$ like a
smooth continuation of $N_{\B}^{\sep}$. There, as soon as component
separation occurs, the bosonic cloud is compressed and the attractive
mean-field due to the boson-boson interaction is amplified to such an
extent that the collapse happens immediately.

\section{Implications for present experiments}
\label{sec:exp}

After investigating the generic properties of degenerate boson-fermion
mixtures in the previous sections I now want to discuss the specific
implications for the mixtures available in present experiments.  In
this section the mass difference and the different oscillator lengths
of the trapping potentials for the two species are included
explicitly.

\subsection{${}^{7}\text{Li}$-${}^{6}\text{Li}$ mixture I}

Quantum degeneracy in a magnetically trapped dilute boson-fermion
mixture was first achieved by Truscott \emph{et al.} \cite{TrSt01}
using a mixture of bosonic ${}^7\text{Li}$ atoms in the $\ket{F=2,
m_{F}=2}$ state and fermionic ${}^6\text{Li}$ with $\ket{F=3/2,
m_{F}=3/2}$. The same mixture was subsequently used by Schreck \emph{et al.}
\cite{ScFe01,ScKh01}. For this particular combination of states the
boson-boson interaction is attractive and the boson-fermion
interaction repulsive with scattering lengths \cite{AbMc97}
\eq{ \label{eq:exp_Li7Li6F2_scattlength}
  a_{\B} = -1.46\,\text{nm} 
  \;,\quad
  a_{\BF} = 2.16\,\text{nm} .
}
The attractive boson-boson interaction can cause the collapse of the
bosonic component and leads to a severe limitation of the number of
bosons (see Sec. \ref{sec:BF+_collapse}). This is a major hindrance
for the implementation of an efficient scheme for the sympathetic
cooling of the fermionic component. Ideally one would like to have a
large bosonic cloud at a very low temperature to act as a coolant for
the fermions. The attractive boson-boson interaction, however,
restricts the number of particles in the Bose-Einstein condensate and
hence sets a lower bound to the temperature of the bosonic cloud. In
conclusion the lowest achievable temperature for the fermionic
component is also restricted.

\begin{figure}
\includegraphics[height=0.26\textheight]{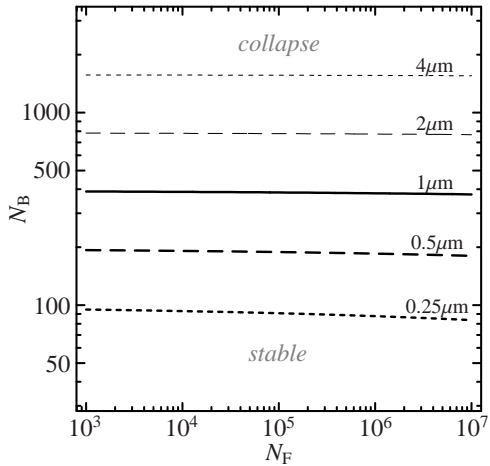}
\caption{Phase diagram for a ${}^7\text{Li}$-${}^6\text{Li}$ mixture
  with scattering lengths \eqref{eq:exp_Li7Li6F2_scattlength}. The
  lines show the stability limit for different values of the
  oscillator length $\ell_{\B}$ of the trap as labeled. For $N_{\B}$
  above the stability limit the bosonic component collapses.}
\label{fig:exp_phasediag_Li7Li6F2}
\end{figure}

The stability limits for this ${}^7\text{Li}$-${}^6\text{Li}$
mixture can be visualized by a phase digram in the plane spanned by
$N_{\F}$ and $N_{\B}$ as shown in
Fig. \ref{fig:exp_phasediag_Li7Li6F2}. Each of the curves marks the
limit of stability for a different size of the external trapping
potential characterized by the oscillator length $\ell_{\B}$ for the
bosonic species. The respective oscillator length $\ell_{\F}$ for the
fermions follows from simple scaling with respect to the different
masses and magnetic moments. In the experiment of Truscott \emph{et
al.} the mean oscillator length is $\ell_{\B}=2.67\,\mu\text{m}$
\cite{TrSt01}, whereas Schreck \emph{et al.} used a trap with
$\ell_{\B}=1.23\,\mu\text{m}$ \cite{ScKh01}. For boson numbers above
the stability limit the bosonic component is unstable against
mean-field collapse. Narrowing of the trapping potential,
i.e. reducing the oscillator length $\ell_{\B}$, enhances the effect
of the interactions.

Obviously the stability of the mixture depends mainly on the boson
number $N_{\B}$ and the critical boson number is to a good
approximation given by the corresponding value of the pure Bose
gas. The fermionic component has essentially no influence on the
stability because the boson-fermion interaction is too weak. This is
in agreement with the experimental findings \cite{TrSt01,ScFe01}. As
discussed in Sec. \ref{sec:BF+_collapse} a stronger boson-fermion
repulsion would destabilize the mixture anyway.

Notice that the calculations presented here are restricted to
isotropic traps. The prolate trap deformation present in most
experiments leads to a reduction of the critical particle number
compared to a spherical trap with same mean oscillator length. For a
pure Bose gas in a trap with $\omega_{z}/\omega_{r} \approx0.1$ as in
the experiment of Truscott \emph{et al.} \cite{TrSt01} the critical
particle number is reduced by typically $20\%$ \cite{GaFr01}. For the
strongly deformed trap with $\omega_z/\omega_r\approx0.02$ used by
Schreck \emph{et al.} \cite{ScKh01} this reduction is in the order of $40\%$.

\subsection{${}^7\text{Li}$-${}^6\text{Li}$ mixture II}

The two Lithium isotopes can also be trapped in another combination of
angular momentum states as was successfully demonstrated by Schreck
\emph{et al.} \cite{ScKh01}. The relevant $s$-wave scattering lengths for
a mixture of ${}^{7}\text{Li}$ in the $\ket{F=1,m_F=-1}$ state and
${}^{6}\text{Li}$ in $\ket{F=1/2,m_F=-1/2}$ are both positive
\cite{AbVe97,ScKh01}
\eq{ \label{eq:exp_Li7Li6F1_scattlength}
  a_{\B} = 0.27\,\text{nm} 
  \;,\quad
  a_{\BF} = 2.01\,\text{nm} .
}
Therefore this mixture is stable with respect to collapse but may
undergo component separation generated by the boson-fermion
repulsion (see Sec. \ref{sec:BF+_separation}).
\begin{figure}
\includegraphics[height=0.26\textheight]{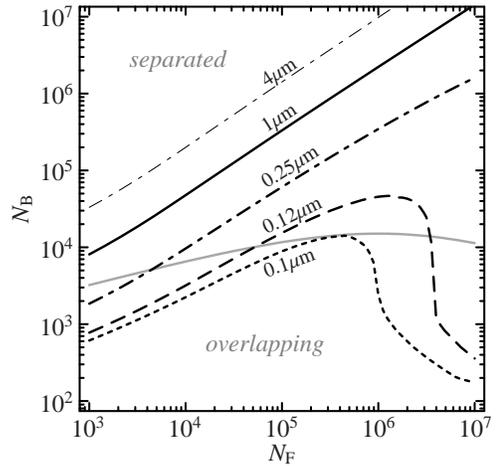}
\caption{Phase diagram for a ${}^7\text{Li}$-${}^6\text{Li}$ mixture
  with scattering lengths \eqref{eq:exp_Li7Li6F1_scattlength}. The
  lines show the onset of component separation for different values of
  the oscillator length $\ell_{\B}$. For $N_{\B}$ above the limit the
  two species separate spatially. The gray curve shows the onset of
  separation for vanishing boson-boson interaction and
  $\ell_{\B}=1\mu\text{m}$.}
\label{fig:exp_phasediag_Li7Li6F1}
\end{figure}

The phase diagram of this mixture in the $N_{\F}$-$N_{\B}$-plane is
shown in Fig. \ref{fig:exp_phasediag_Li7Li6F1}. For values of $N_{\B}$
above the phase boundaries plotted for different oscillator lengths
the mixture separates spatially. The onset of separation for the
scattering lengths \eqref{eq:exp_Li7Li6F1_scattlength} depends in a
nontrivial way on the particle numbers and the oscillator
length. Especially for tightly confining traps with
$\ell_{\B}<0.15\,\mu\text{m}$ the phase boundary at large $N_{\F}$
suddenly bends towards smaller boson numbers. As discussed in
Sec. \ref{sec:BF+_separation} this structure is associated with a
change in the order of the phase transition. At small $N_{\F}$ or for
shallow traps the transition is continuous, i.e. the central fermion
density continuously goes to zero as $N_{\B}$ is increased. For large
$N_{\F}$ and tightly confining traps the transition is of first order,
that is the central fermion density jumps to zero if the phase
boundary is crossed. This effect is solely caused by the presence of
the weak boson-boson repulsion.

We should like to point out that although the boson-boson scattering
length is by one order of magnitude smaller than the boson-fermion
scattering length it has a strong influence on the phase diagram also
for shallow traps and small particle numbers. The gray curve in Fig.
\ref{fig:exp_phasediag_Li7Li6F1} shows the phase boundary for
$\ell_{\B}=1\mu\text{m}$ without the boson-boson interaction (compare
with the black solid curve). The boson-boson repulsion stabilizes the
system substantially, i.e. the onset of separation is shifted to much
larger particle numbers.

Schreck \emph{et al.} \cite{ScKh01} managed to produce such a mixture
with $N_{\B}\approx10^4$ and $N_{\F}\approx 2.5\times 10^4$ in a trap
with mean oscillator length of $\ell_{\B}=1.0\,\mu\text{m}$ (solid curve
in Fig \ref{fig:exp_phasediag_Li7Li6F1}). According to our
calculations (effects of the trap anisotropy are not included) the
bosonic and fermionic density are still overlapping which is consistent
with the experimental findings. However, an increase of the boson number
by a factor $10$ or a reduction of the fermion number to
$N_{\F}\approx 10^3$ would already lead into the regime of separation.

\subsection{${}^{87}\text{Rb}$-${}^{40}\text{K}$ mixture}
\label{sec:exp_Rb87K40}

A first step towards an ultracold mixture of ${}^{87}\text{Rb}$ and the
fermionic ${}^{40}\text{K}$ was reported by Goldwin \emph{et al.}
\cite{GoPa02} and recently Roati \emph{et al.} \cite{RoRi02} managed to
reach quantum degeneracy. This system is particularly interesting, because
it shows a large negative boson-fermion scattering length \cite{FeIn02}
\eq{ \label{eq:exp_Rb87K40_scattlength}
  a_{\B} = 5.25\,\text{nm} 
  \;,\quad
  a_{\BF} = -13.8\,\text{nm} .
}
Thus this mixture can undergo a simultaneous collapse of the bosonic
and fermionic density distribution as discussed in
Sec. \ref{sec:BF-_collapse}.
\begin{figure}
\includegraphics[height=0.26\textheight]{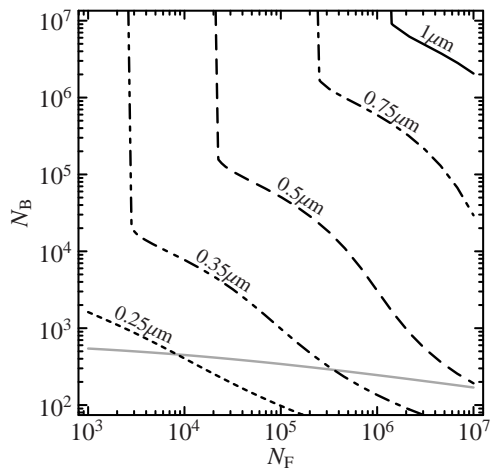}
\caption{Phase diagram for the ${}^{87}\text{Rb}$-${}^{40}\text{K}$ mixture
  with scattering lengths \eqref{eq:exp_Rb87K40_scattlength}. The
  lines show the stability limit for different values of the
  oscillator length $\ell_{\B}$. For $N_{\B}$ above the limit the two
  species collapse simultaneously. The gray curve shows the stability
  limit for vanishing boson-boson interaction with
  $\ell_{\B}=1\mu\text{m}$.}
\label{fig:exp_phasediag_Rb87K40}
\end{figure}

Figure \ref{fig:exp_phasediag_Rb87K40} shows the corresponding phase
diagram in the $N_{\F}$-$N_{\B}$-plane. One might expect that the
strong boson-fermion attraction sets a very severe limit to the size
of a stable mixture. In principle this expectation is correct as
illustrated by the gray solid curve which corresponds to the mere
boson-fermion attraction for $\ell_{\B}=1\,\mu\text{m}$. Without the
boson-boson interaction the mixture would collapse if the boson number
exceeds $N_{\B}^{\crit}\approx300$.

Fortunately a moderate repulsive boson-boson interaction is present
which has a tremendous influence on the phase diagram. First of all
the critical boson numbers are increased by roughly four orders of
magnitude. Moreover, below a certain fermion number marked by the
vertical pieces of the curves in Fig. \ref{fig:exp_phasediag_Rb87K40}
the mixture is stable up to arbitrary boson numbers; this is the
effect of absolute stabilization due to the boson-boson repulsion
discussed in Sec. \ref{sec:BF-_collapse}. Thus in a trap with
$\ell_{\B}=1\,\mu\text{m}$ any combination of boson and fermion numbers
up to $N_{\B}\approx N_{\F}\approx 10^7$ would be stable although a
strongly attractive boson-fermion interaction is present.

The particular combination of scattering lengths of the
${}^{87}\text{Rb}$-${}^{40}\text{K}$ mixture opens interesting
perspectives for the implementation of a sympathetic cooling scheme. The
strong boson-fermion attraction generates an enhancement of the boson
and fermion density in the overlap region which is --- due to the
boson-boson repulsion --- expanded over a large volume (see
Fig. \ref{fig:BF-_densities_varBF_2}). Therefore a large fraction of
the fermionic cloud overlaps with the bosonic distribution and enables
a very efficient inter-species thermalization. At the same time the
boson-boson repulsion stabilizes the system against simultaneous
collapse for all experimentally relevant particle numbers.

\subsection{${}^{23}\text{Na}$-${}^{6}\text{Li}$ mixture}

Recently, Hadzibabic \emph{et al.} \cite{HaSt01} reported the sympathetic
cooling of a mixture of ${}^{23}\text{Na}$ and ${}^{6}\text{Li}$ to
quantum degeneracy. To my knowledge there is no data on the inter-species
scattering length for this system yet. However, since a stable mixture
with large numbers of bosons and fermions ($N_{\B}\approx2\times 10^6$,
$N_{\F}=\approx1.5\times 10^5$) was produced one can at least deduce
bounds for the boson-fermion scattering length. Assuming $a_{\B} =
2.75\,\text{nm}$ \cite{InAn98} one finds that the boson-fermion scattering
length must obey $a_{\BF} > -14.6\,\text{nm}$ in order to allow a stable
mixture in this experiment. Moreover, the linear density profiles of the
fermionic cloud shown in \cite{HaSt01} suggest that the components are not
separated. From the minimal requirement that the fermion density in the
center is larger than zero one obtains an upper bound
$a_{\BF}<5.1\,\text{nm}$ \footnote{The requirement that the fermionic
density profile drops monotonically, i.e. does not show a central dip,
leads to the more restrictive upper bound $a_{\BF}<0.69\,\text{nm}$.}.
Notice that these estimates do not account for the effects of the trap
anisotropy and the finite temperature present in the experiment.

\section{Conclusions}

In summary, I have presented a systematic study of the structure and
stability of trapped binary boson-fermion mixtures at zero
temperature. The bosonic component is described by a modified
Gross-Pitaevskii equation which includes the mean-field interaction
with the fermionic component. The fermionic density distribution is
obtained within the Thomas-Fermi approximation.

Depending on the strengths of the boson-boson and the boson-fermion
interactions, characterized by the $s$-wave scattering lengths $a_{\B}$
and $a_{\BF}$, respectively, different structural phenomena can be
observed:
\begin{description}
\item[\textmd{$a_{\B}<0$}]: The collapse of the bosonic component
dominates the phase diagram and sets a severe limitation to the
maximum number of bosons in a stable mixture. The boson-fermion
interaction has only small influence, however, attractive as well as
repulsive boson-fermion interactions promote the collapse of the
bosonic component.

\item[\textmd{$a_{\B}\ge0,\, a_{\BF}<0$}]: The boson-fermion
attraction enhances the density of both species in the overlap region
and can cause a simultaneous collapse of the bosonic and the fermionic
cloud. A repulsive boson-boson interaction stabilizes the system, i.e.
increases the critical particle numbers.

\item[\textmd{$a_{\B}\ge0,\, a_{\BF}>0$}]: Repulsive boson-fermion
interactions reduce the overlap of the two species and eventually lead to
the spatial separation of the components.  Again a repulsive boson-boson
interaction stabilizes the mixture against separation. \end{description}
The rich variety of structures generated by the interplay of boson-boson
and boson-fermion interactions has important consequences for the
experimental realization of sympathetic cooling schemes. As far as the
density distributions are concerned efficient sympathetic cooling relies
on (a) the mechanical stability of the mixture and (b) a large overlap of
the two species.

From this point of view mixtures with repulsive boson-boson
interactions ($a_{\B}>0$) are superior since they do not suffer the
severe limitation of the boson number imposed by the collapse for
$a_{\B}<0$. The boson-fermion cross section and thus the modulus of
the boson-fermion scattering length should be sufficiently large to
enable efficient inter-species thermalization. For $a_{\BF}>0$ this
may lead into problems with the separation of the two components, for
$a_{\BF}<0$ there may be a simultaneous collapse of both
clouds. Fortunately, a moderate boson-boson repulsion can stabilize
the mixture against both transitions.

Regarding the structure of the density profiles a mixture with
$a_{\B}>0$ and $a_{\BF}<0$ --- like the
${}^{87}\text{Rb}$-${}^{40}\text{K}$ mixture discussed in Sec.
\ref{sec:exp_Rb87K40} --- is especially appealing for the
implementation of an efficient sympathetic cooling scheme. It is
stabilized against collapse by the boson-boson repulsion and exhibits
a large overlap between the species due to an extended boson
distribution and a significantly increased fermion density within the
overlap volume.

\section*{Acknowledgments}

I would like to thank H. Feldmeier and K. Burnett for stimulating
discussions. This work was supported by the Deutsche
Forschungsgemeinschaft (DFG).
\vfil



\end{document}